\documentclass[preprint,showpacs,aps,draft]{revtex4}
\usepackage{bm}

\begin{document}

\title{Multiscaling of galactic cosmic ray flux}

\author{A. Bershadskii}
\affiliation{ICAR, P.O. Box 31155, Jerusalem 91000, Israel}

\begin{abstract}
Multiscaling analysis of differential flux dissipation rate of galactic cosmic rays (Carbon nuclei) 
is performed in the energy ranges:  56.3-73.4 Mev/nucleon and 183.1-198.7 MeV/nucleon, 
using the data collected by ACE/CRIS spacecraft instrument for 2000 year. 
The analysis reveals strong (turbulence-like) intermittency of the flux dissipation rate for the 
short-term intervals: 1-30 hours. It is also found that type of the intermittency 
can be different in different energy ranges.  
\end{abstract}


\maketitle
\section{Introduction}

  Galactic cosmic rays (GCR) originate outside the solar system. 
They comprise protons (85$\%$), alpha particles (14$\%$) and heavy nuclei. 
Interaction with large-scale 
ordered magnetic field causes the gradient- and curvature-drift motion of GCR in the inner 
heliosphere, while the interaction with the irregular (stochastic) field component results 
in the pitch angle scattering of GCR. 
The scale sizes for the two effects, drifts 
which depend on variations in the mean field on the order of the heliocentric radial 
distance, and diffusion which depends on irregularities of sizes comparable to the particles 
gyro-radii, are quite distinct (see, for instance \cite{fs},\cite{alania} and references therein). 

  Until recently space instruments have lacked the combination of large geometrical 
factor and good mass resolution required to address the question of the variation on 
{\it short} time scales of individual nuclides other than $H$ and $He$. In present 
Letter we report on an investigation of the flux fluctuations of GCR nuclei 
(Carbon - $C$: $z=6$) based on the data obtained by instrumentation carried aboard the Advanced 
Composition Explorer (ACE) spacecraft for 2000 year. The intensities of these 
low energy particles have been continuously monitored by the Cosmic Ray Isotope 
Spectrometr (CRIS). 

In order to get away from the effects of the Earth's magnetic field, the ACE spacecraft orbits 
at the L1 libration point which is a point of Earth-Sun gravitational 
equilibrium about 1.5 million km from Earth and 148.5 million km from the Sun. 

 The large collecting power and high 
resolution of the CRIS instrument allow us to investigate  modulation 
of GCR flux on short time scales (beginning from 1 hour) and in different energy ranges.

 Current theories of parallel and perpendicular diffusion are fairly well accepted for high 
energy particles. 
However, there are important outstanding issues pertaining to diffusion and transport of 
charged particles at medium to low energies. The particles at low energies are 
believed interact directly with the steepened dissipation range that sets in near 
the ion inertial scale. Dynamical scattering theory suggests that these particles may be 
sensitive to (1) the dynamical decorrelation mechanism that operates at these scales; (2) 
the "geometry" of the fluctuations in the dissipation range; and (3) the spectral 
distribution of power at the small scales \cite{matt1}.        \\

  Multiscaling properties of cosmic rays have been already discussed in the light of 
the data obtained by neutron monitors  with cut-off rigity 1.09 GeV 
(see, for instance, \cite{kv},\cite{ber1} and references therein) as well as modulation 
of the cosmic rays by solar wind turbulence and magnetic field 
intermittency \cite{hor1}-\cite{hor2}. However, 
as far as we know, multiscaling analysis of the GCR heavy nuclei data obtained by satellite 
instruments is performed here for the first time. This analysis reveals a strong  (turbulent-like) 
intermittency of the GCR dissipation rate and dependence of the intermittency type from the 
energy ranges (not to be confused with the HEP multiplicity intermittency \cite{bp},\cite{ber2}).

\section{Differential flux dissipation rate}

  Differential flux of the GCR: $J$, as measured by CRIS, is proportional to the normal 
component of the velocity field of the GCR: $u$, and to their local concentarion:  
$n$,
$$
J \sim un    \eqno{(1)}
$$
Fluctuations of $J$ with time are then determined by corresponding 
fluctuations of $u$ and $n$. 

Dissipation of passive admixture concentration in fluid turbulence is 
characterized by a "gradient" measure \cite{my}-\cite{sa}:
$$
\chi_r =\frac{\int_{v_r} (\bigtriangledown{n})^2 dv}{v_r}    \eqno{(2)}
$$
where $v_r$ is a subvolume with space-scale $r$ (for detail justification of 
this measure see handbook \cite{my}: p. 381 and further). Scaling law of this 
measure moments,
$$
\langle \chi_{r}^p \rangle \sim r^{-\mu_p}      \eqno{(3)}
$$
is an important characteristic of the dissipation rate field just in {\it inertial} 
interval of turbulence  (see, for instance, \cite{my}-\cite{sa}). Analogous measure 
is used to characterize also dissipation rate of turbulent velocity (or kinetic energy) 
in the inertial interval of scales \cite{my}-\cite{sa}. 

For turbulent flows the Taylor hypothesis is generally used to interpret 
the data \cite{hor1},\cite{hor2},\cite{s}. 
This hypothesis states that the intrinsic time dependence of the wavefields ($u$ and $n$) 
can be ignored when the turbulence is convected past the probes at nearly constant 
speed. With this hypothesis, the temporal dynamics should reflect the spatial one, 
i.e. the fluctuating velocity (concentration) field measured by a given probe 
as a function of time, $u(t)$ is the same as the velocity $u(x/\langle u \rangle)$ 
where $\langle u \rangle$ is the mean velocity and $x$ is the distance to a position 
"upstream" where the velocity is measured at $t=0$. 

  With the Taylor hypothesis $dJ/dx$ is replaced by $dJ/\langle u \rangle dt$ and 
one can define GCR flux dissipation rate as:
$$
\chi_{\tau} \sim \frac{\int_0^{\tau} (\frac{dJ}{dt})^2 dt}{\tau}    
\eqno{(4)}
$$
where $\tau \simeq r/\langle u \rangle$ and corresponding scaling of 
the dissipation rate moments as \cite{sa}:
$$
\langle \chi_{\tau}^p \rangle \sim \tau^{-\mu_p}      \eqno{(5)}
$$
     
 Substituting (1) into (4) one can estimate the GCR flux 
dissipation rate through characteristics of $u$ and $n$. We can consider 
two asymptotic regimes. For sufficiently small GCR energies the flux 
dissipation rate has been dominated by the concentration dissipation:
$$
\langle \chi_{\tau}^p \rangle \sim \langle [\frac{1}{\tau}\int_0^{\tau} 
(\frac{dn}{dt})^2 dt]^p \rangle \sim \tau^{-\mu_p}   
\eqno{(6)}
$$
while for sufficiently large GCR energies the flux dissipation rate has been 
dominated by the energy (velocity) dissipation:
$$
\langle \chi_{\tau}^p \rangle \sim \langle [\frac{1}{\tau}\int_0^{\tau} 
(\frac{du}{dt})^2 dt]^p \rangle \sim \tau^{-\mu_p}   
\eqno{(7)}
$$ 

\section{The data}

 As it was mentioned in Introduction we will use the data collected 
by ACE/CRIS instrument during 2000 year (the year of maximum solar 
activity). We will consider carbon nuclei $C$ ($z=6$), due to carbon 
is the lightest abundant nuclei (after $H$ and $He$) in the ACE/CRIS 
collection. As we shall see this fact allow us consider the both asymptotes
mentioned in previous Section. 
  
  Figure 1a shows scaling of the GCR flux dissipation rate moments 
$\langle \chi_{\tau}^p \rangle$ (4)-(5) for the C-nuclei with energies 
from energy range: 56.3-73.4 Mev/nucleon (the lowest energy range 
in the ACE/CRIS collection for carbon). 

Figure 1b shows the scaling exponents $\mu_p$ (circles) extracted from figure 1a 
(as slopes of the straight lines, cf (5)). The solid curve in figure 1b corresponds to the 
intermittency exponents $\mu_p$ obtained for the {\it inertial}-convective region of 
a passive admixture concentration in a laboratory turbulent air flow \cite{sa}.  
To support the striking correspondence between the two data sets multiscaling let us 
calculate also the Extended Self-Similarity (ESS \cite{ben}) exponents, 
$\beta_p$, extracted from equation:
$$
\langle \chi_{\tau}^p \rangle \sim \langle \chi_{\tau}^3 \rangle^{\beta_p} 
\eqno{(8)}
$$
The ESS of type (8) usually has clearer scaling form than ordinary scaling (5) 
and covers a wider range of scales (see \cite{ben} for a review of ESS and for 
examples). 
Figure 2a shows the ESS of the GCR flux dissipation rate moments 
$\langle \chi_{\tau}^p \rangle$ (8) using log-log scales. The straight 
lines (the best fit) are drawn to indicate the ESS (8). Figure 2b shows the ESS 
exponents $\beta_p$ (circles) extracted from figure 2a 
(as slopes of the straight lines, cf (8)). The solid curve in figure 2b corresponds to the 
intermittency exponents $\beta_p$ obtained for the {\it inertial}-convective region of 
passive admixture concentration in different laboratory turbulent flows 
\cite{sa}.  

  Now let us turn to the data corresponding to energy range with the highest 
for the carbon nuclei energies observed by ACE/CRIS.  
Figure 3a shows scaling of the GCR flux 
dissipation rate moments 
$\langle \chi_{\tau}^p \rangle$ (4)-(5) for the C-nuclei with energies 
from energy range: 183.1-198.7 MeV/nucleon. 

Figure 3b shows the scaling exponents $\mu_p$ (circles) extracted from figure 3a 
(as slopes of the straight lines, cf (5)). The solid curve in figure 3b 
corresponds to the intermittency exponents $\mu_p$ calculated using 
the She-Leveque model \cite{sl}, which is in very good agreement with the data 
for {\it velocity} (kinetic energy dissipation) field intermittency obtained in {\it inertial} interval 
for isotropic fluid turbulence. 
Figures 4a and 4b shows corresponding Extended 
Self-Similarity (ESS) properties
$$
\langle \chi_{\tau}^p \rangle \sim \langle \chi_{\tau}^4 \rangle^{\beta_p} 
\eqno{(9)}
$$
observed for the data. Again, as in figure 3b, correspondence to the fluid turbulence 
(energy dissipation rate - the solid curve in figure 4b) is very good. 

The interval of time scales under consideration is 1-30 hours. It seems 
to be useful to compare the observed properties of GCR flux dissipation rate
with relevant properties of the local interplanetary magnetic field. 
Figure 5 shows energy spectrum of 3D magnetic field measured by ACE/MAG 
magnetometer in the same time intervals 
as the ACE/CRIS data. The slope -5/3 indicates Kolmogorov-like scaling (cf \cite{gold}). Let us 
recall that the turbulent fluid dissipation rates used for comparison in figures 1-4 
were obtained just for inertial (inertial-convection) interval of scales where 
the Kolmogorov-like scaling should to be expected \cite{sa},\cite{sl}.             

\section{Discussion}
   The results of previous Section seems to be in agreement with the 
qualitative estimates made in Section 2: for relatively small energies the flux dissipation 
rate behaves as one dominated by the GCR particles concentration dissipation rate, whereas 
for relatively large energies the GCR flux dissipation rate seems to be dominated 
by velocity (kinetic energy) dissipation. 

  Since CRIS instrument measures the differential flux ($J$) only, but not 
the velocity of GCR ($u$) and their concentration ($n$) separately (cf (1)) 
we have no idea about magnitudes of the fluctuations of $u$ and $n$ 
themselves. Therefore existence of the two asymptotes (6) and (7) is still 
a pure phenomenology (for the intermediate energies the scaling is deformed 
by competition between the two different mechanisms of the flux dissipation). 
The same reason (unknown $u$ and $n$ for the GCR) makes it impossible to 
perform any theoretical estimates for the scales of the GCR flux 
dissipation rate multiscaling.

  Although correspondence between the motions in MHD and hydro cases has 
been reported earlier (see, for instance, very recent paper \cite{cho2} and references 
therein) it is much more difficult to understand such detail correspondence between 
cosmic rays and fluid turbulence. Indeed, as it follows from 
previous two sections the observed GCR particles dissipation rate exhibits 
intermittent properties of a passive admixture convected by turbulent motion of a classic 
(isotropic, incompressible) non-magnetic fluid. While the solar wind (including 
the interplanetary magnetic field - IMF) is certainly very different from such a fluid. 
Moreover, 
interaction of the electrically charged GCR particles with the solar wind 
and the magnetic field is expected to have strong resonance features (see, 
for instance, \cite{chan}-\cite{clv},\cite{cho2} and references therein).  Therefore, even 
the fact that many characteristics of solar wind and IMF turbulence are known to be 
similar to those of turbulence of classic non-magnetic fluids 
(see \cite{hor1}-\cite{hor2} and references therein) the presumably 
{\it resonance}-like interactions of GCR particles with solar wind and IMF seems to be non-consistent with 
the observed here picture, that request reconsideration of our approach to stochastic {\it short}-term 
convection of the low-energy GCR particles (heavy nuclei) in the heliosphere and  their {\it short}-term 
interactions with solar wind and IMF.  This reconsideration represents a serious challenge 
for a wide scientific community: both for astrophysicists and for plasma 
experts.   \\

  The author is grateful to ACE/CRIS and to ACE/MAG instrument teams as 
well as to the ACE Science Center for providing the data.  Numerous discussions on the subject with K.R. Sreenivasan as well as the Referees' comments and suggestions were very useful for this investigation.

\newpage

\centerline{\bf Figure Captions}

Figure 1a.  
The GCR flux dissipation rate moments $\langle \chi_{\tau}^p \rangle$ against $\tau$ 
for the C-nuclei with energies from energy range: 56.3-73.4 Mev/nucleon. 
The time interval $\tau$ is 
measured in hours, whereas the GCR differential flux $J$ is measured 
in $10^{-7.5} ~particles/m^2~s~sr~Mev$. Log-log scales are chosen in 
the figure for comparison with scaling equation (5). The straight 
lines (the best fit) are drawn to indicate the scaling.  The upper data sets 
correspond to larger $p=2,3,4,5$\\

Figure 1b. 
The scaling exponents $\mu_p$ (circles) extracted from figure 1a. 
The solid curve corresponds to the 
intermittency exponents $\mu_p$ obtained for the inertial-convective region of 
a passive admixture concentration dissipation rate in a laboratory turbulent air flow \cite{sa}.  \\

Figure 2a. The ESS of the GCR flux dissipation rate moments 
$\langle \chi_{\tau}^p \rangle$ against $\langle \chi_{\tau}^3\rangle$ 
in log-log scales (8) for energy range: 56.3-73.4 Mev/nucleon. 
The straight lines (the best fit) are drawn to indicate the ESS (8). The upper data sets 
correspond to larger $p=2,3,4,5$. \\

Figure 2b. The ESS exponents $\beta_p$ (circles) extracted from figure 2a. 
The solid curve corresponds to the 
intermittency exponents $\beta_p$ obtained for the inertial-convective region of 
passive admixture concentration dissipation rate in different fluid turbulent flows 
\cite{sa}.  \\

Figure 3a.   The GCR flux dissipation rate moments 
$\langle \chi_{\tau}^p \rangle$ (4)-(5) against $\tau$ for the C-nuclei with energies 
from energy range: 183.1-198.7 MeV/nucleon. \\

Figure 3b. The scaling exponents $\mu_p$ (circles) extracted from figure 3a. 
The solid curve in figure 3b 
corresponds to the intermittency exponents $\mu_p$ calculated using 
the She-Leveque model for isotropic fluid turbulence \cite{sl}. \\

Figure 4a. The ESS of the GCR flux dissipation rate moments 
$\langle \chi_{\tau}^p \rangle$ against $\langle \chi_{\tau}^3\rangle$ in log-log scales (9) 
for energy range: 183.1-198.7 MeV/nucleon. 
The straight lines (the best fit) are drawn to indicate the ESS (9). \\

Figure 4b. The ESS exponents $\beta_p$ (circles) extracted from figure 4a. 
The solid curve corresponds to the 
intermittency exponents $\beta_p$ calculated using the She-Leveque model. \\

Figure 5. Energy spectrum of 3D magnetic field measured by ACE/MAG magnetometer 
in the discussed range of the time-scales.  

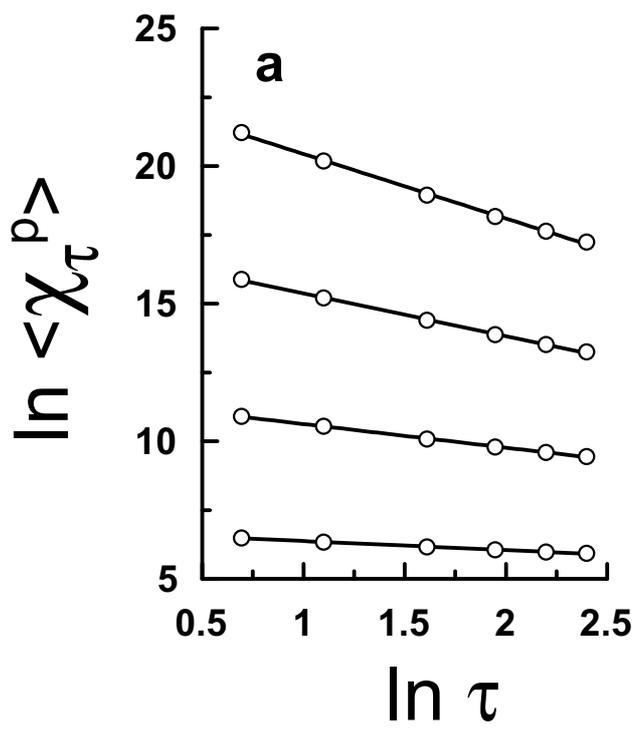
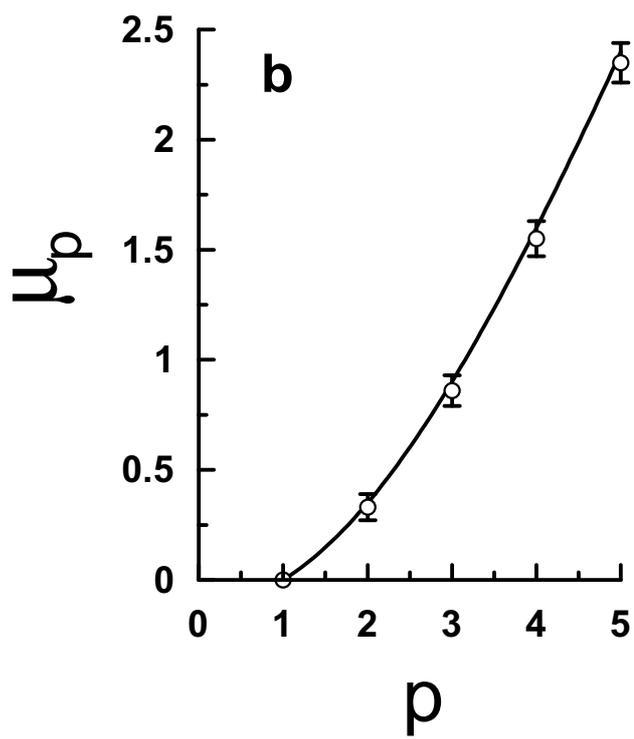
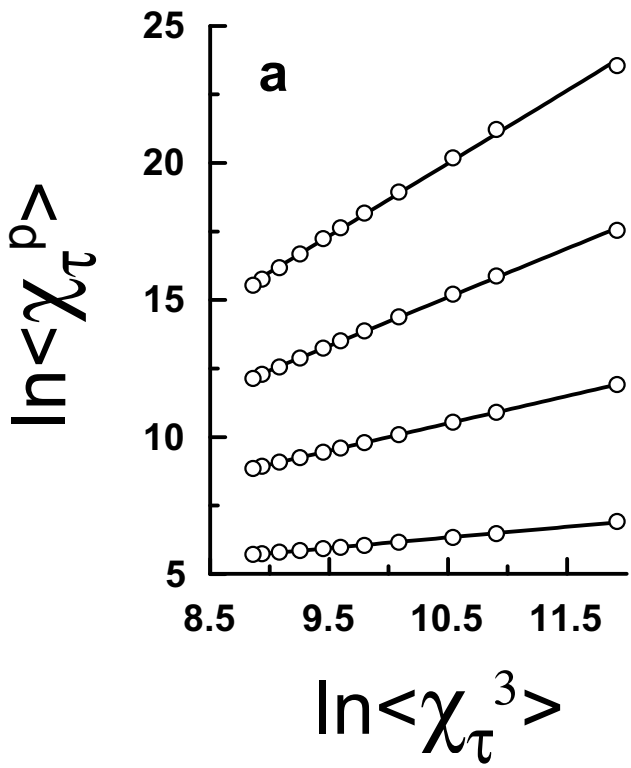
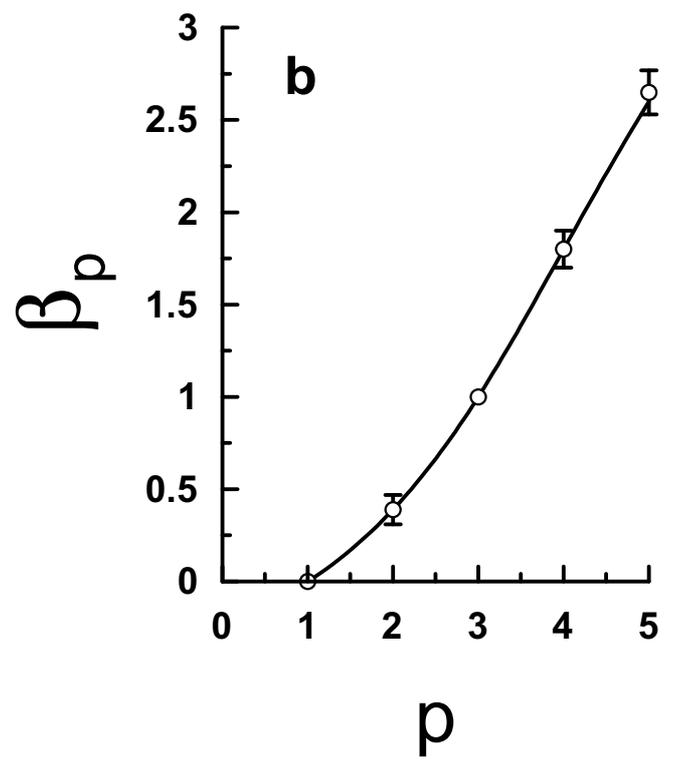
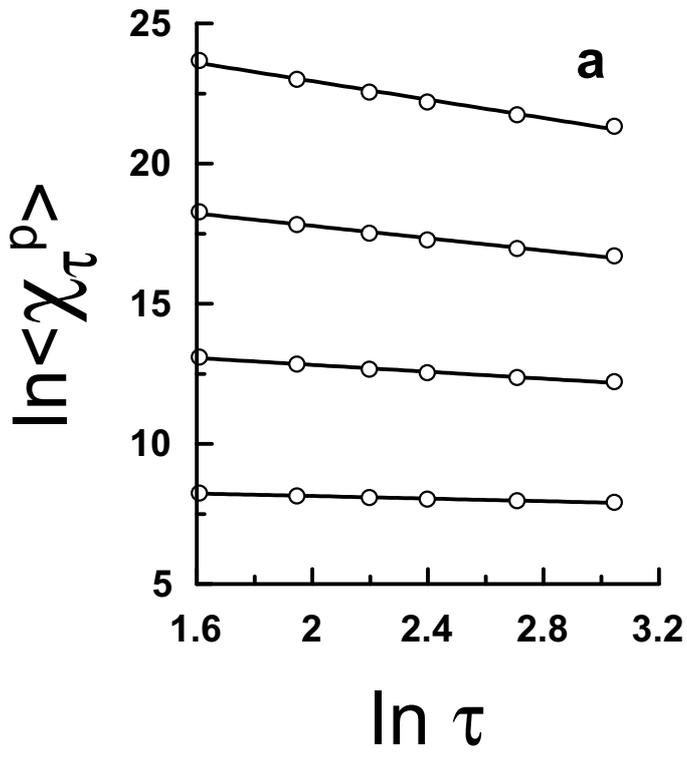
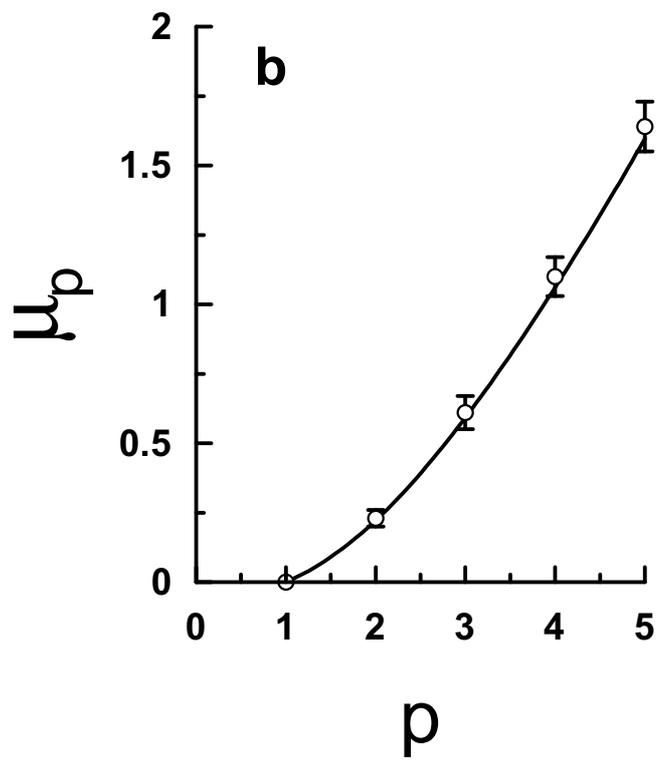
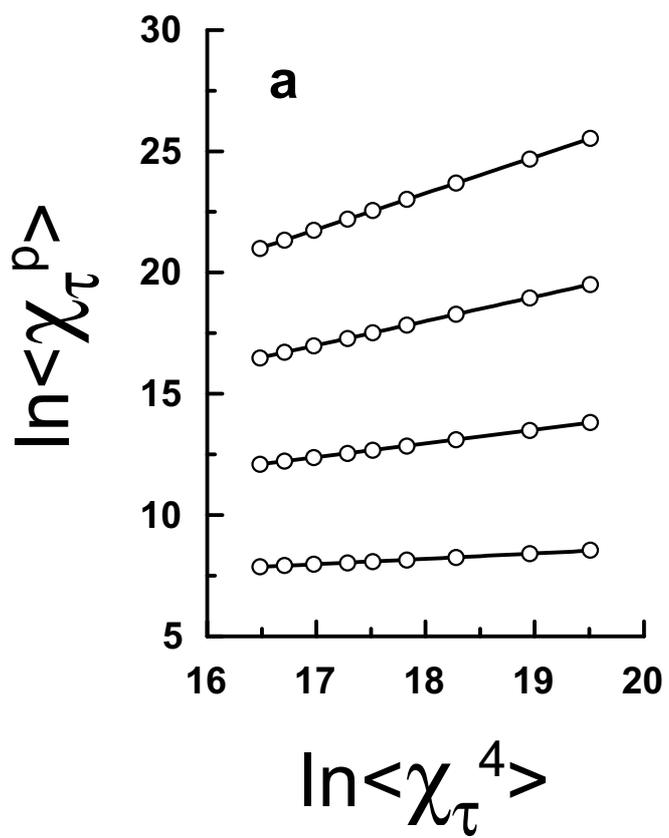
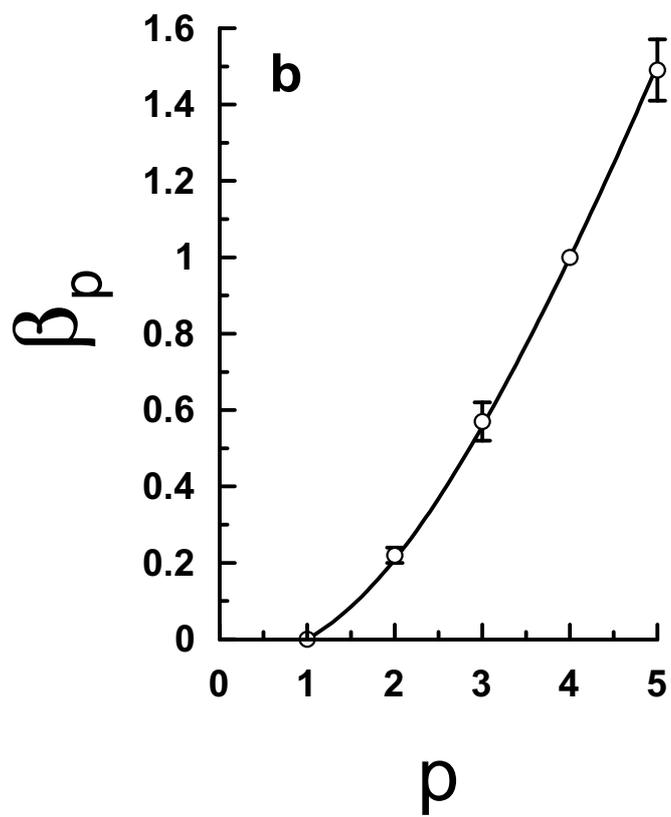
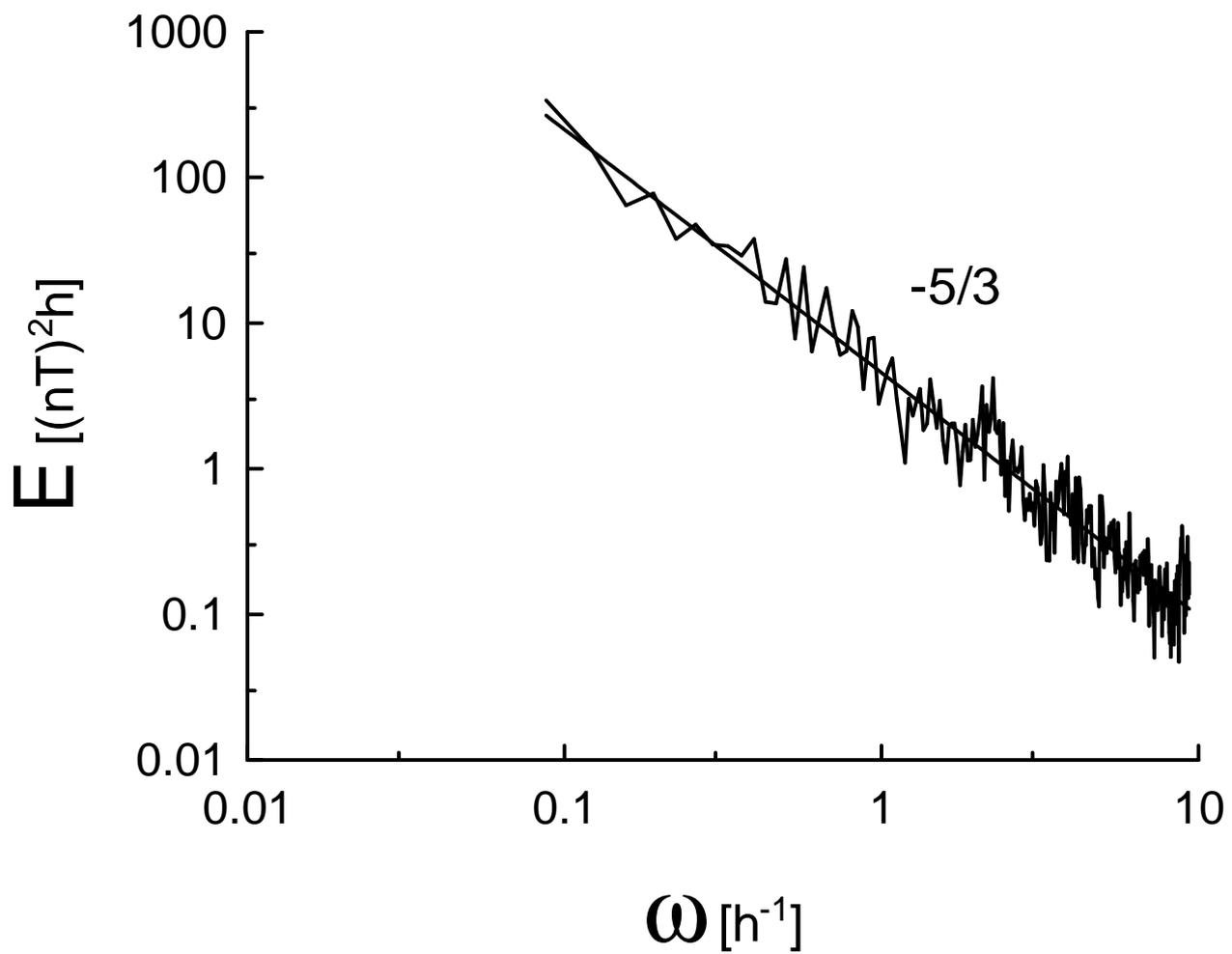
\end{document}